\documentclass[12pt]{article}
\usepackage[margin=1in]{geometry}
\usepackage{indentfirst}
\usepackage{graphicx}
\usepackage{amsmath}
\newcommand{\nocomma}{}
\newcommand{\tmem}[1]{{\em #1\/}}
\newcommand{\tmtextbf}[1]{{\bfseries{#1}}}
\newcommand{\tmtextit}[1]{{\itshape{#1}}}
\newcommand{\tmtexttt}[1]{{\ttfamily{#1}}}
\newenvironment{itemizedot}{\begin{itemize} }{\end{itemize}}
\begin{document}

\title{Temperature, Abundance, and Mass Density Profiling of the Perseus Galaxy Cluster}\author{Paul Geringer }\maketitle

\begin{abstract}
  Clusters of galaxies are massive structures containing approximately $10^{15}$ solar masses. The mass contained in clusters provided the first hints of the existence of dark matter. The hot intra-cluster medium (ICM), which is a gas with a temperature of $10^{8}$ Kelvin, is easily detected by X-Ray telescopes. Of particular interest is the close Perseus galaxy cluster, one of the brightest X-Ray emitting galaxy clusters in the sky. Due to the large angular extent of the cluster, background noise subtraction is particularly challenging, and has yet to be completed using the full set of Perseus observations from the XMM-Newton X-Ray Telescope. Detailed temperature and abundance radial profile maps have revealed a significant lack of homogeneity within the cluster. Previous surveys of Perseus with the Suzaku telescope, which has a worse angular resolution and less light collecting area than XMM-Newton, revealed over-densities of X-Ray emission. These results provide evidence that the baryon fraction exceeds the universal average, which we had initially hoped to study. We have yet to confirm or deny the existence of clumping in these regions, which could explain such over-abundance of X-Ray emission. 
  
  \indent{}This project offers a framework of efficient,
  automated processing techniques to ``clean'' images of noise from
  the mechanics of the telescope, background radiation from local sources such
  as the solar wind, and more distant sources such as background AGN. The
  galaxy cluster studied in this project contains high levels of contamination
  due to its line-of-sight position close to the dust- and
  star-filled arms of the Milky Way galaxy. Rigorous spectral model fitting of
  the cluster employ multiple parameters dedicated to accounting for
  these contaminations. The framework created from this analysis technique
  will provide the opportunity to expand this analysis to any nearby galaxy cluster, such as the Virgo, Coma, and Ophiuchus Clusters.
  This research should provide significant insight into how matter, both
  baryonic and dark matter, is distributed throughout diffuse cluster systems, as well as give clues to the origin of the ICM.
\end{abstract}

\clearpage

{\tableofcontents}

\clearpage

\section{Introduction}
\subsection{Galaxy Clusters}\label{galclus}
\paragraph{}The study of X-Ray images of galaxy clusters is a relatively new field of astrophysical study that has benefitted greatly in the past decade from the launch of space based X-Ray observatories. Galaxy clusters are large, gravitationally bound systems consisting of many galaxies. The study of the physical characteristics of galaxy clusters offers key insight into cosmological parameters, such as the distribution of dark matter in the universe, the effects of the cosmological constant, and the distribution of baryonic matter in these enormous structures. Clusters often contain a central active galactic nuclei (AGN), which is theoretically powered by a matter-accreting supermassive black hole. Clusters themselves can be active as well; recently having experienced a collision with another cluster or adding a new component galaxy could be a potential cause for the cluster ICM to not be in hydrostatic equilibrium. Clusters that are in a relaxed hydrostatic equilibrium (\ref{hydro}) state are typically modeled as spherically symmetric systems. 
\begin{equation}
\frac{d\rho}{dr} =  - \rho_g(r)\frac{GM(r)}{r^2}
\label{hydro}
\end{equation}
Clusters offer numerous potential avenues of study. Clusters are used as model laboratories for investigating large scale creation and evolution of elements, as the large amounts of interacting gas are theorized to originate from stellar fusion processes. The extremely thin quantity of hot gas ($< 10^{-3}\ $particles/cm$^{-3}$) in the ICM has emissions well within the X-Ray wavelength range. The composition and exact origin of this gas is an active area of research, which this study aims to explore in depth.
\subsection{Cluster Composition}\label{cluscomp}
\paragraph{}The main elemental composition of the ICM is important for a variety of reasons. The exact abundance of specific elements, such as iron, provides researchers with the ability to model the age and necessary mass of progenitor stars. Iron, which is produced as the upper fusion limit in high-mass stellar cores, can also be produced by the violent explosion of white dwarf stars in Type Ia supernovae. Mergers of the cluster with new component galaxies infuse the colliding ICM with high levels of supernova ejecta. By measuring the distribution of iron abundances, insight can be made into the evolutionary past of the star forming regions of the component galaxies of the cluster \cite{arnaud_temperature_1994}. 
Similarly, alpha element abundances, which consist of O, Mg, Si, S, Ca and Ti, are of great importance to the study of the origin of the first stars and supernova. Following the more recent evolutionary track and formation of elements provides a more complete picture into how the current elemental composition of the universe came to be \cite{arnett_first_1999}. Recent studies have indicated that the majority of the iron abundance observed is directly formed in Type II supernovae, models of which are used to provide constraints on the rate of present and past star formation. This places limits on the timescale during which the iron and alpha element enrichment could have occurred \cite{mushotzky_lack_1997}.
\subsection{Mass Determination}
\paragraph{}A frequently used technique in mass determinations is to assume the ICM of the cluster in question exhibits hydrostatic equilibrium as well as spherical symmetry. Using the model of hydrostatic equilibrium, one can derive a homogenous fit of gravitating mass across the observed portions of the cluster \cite{reiprich_mass_2002}. Such mass determinations have previously been emission model-dependent, relying upon multiple \emph{a priori} assumptions concerning the state of the cluster beyond the aforementioned hypothesis of hydrostatic equilibrium and spherical symmetry. Recent advancements have been made to eliminate some of the \emph{a priori} assumptions made in the case of mass determinations. In particular, the Navarro, Frenk, and White mass density model revealed that the density of dark matter halos around galaxy clusters, which form during the relaxation stage following mergers and collisions, do not necessarily depend on cosmological constants ($\Lambda$ and $\Omega_m$) or the power spectrum fit applied to the dark matter halo  \cite{navarro_universal_1996}.
In recent years, even more significant advancements have been made, culminating in the development of cluster mass mixing models for X-Ray observations. In particular, the \texttt{\texttt{CLMASS}} and \texttt{\texttt{NFWMASS}} models provide a robust method of producing density and gravitational potentials for the spherical shells of clusters in hydrostatic equilibrium.  In particular, when high quality observational data is available, these models provide significant improvements over the classical approach to mass determination \cite{nulsen_model-independent_2010}.
\subsection{Motivation for Study}	
\paragraph{}This study concerns the analysis of archival data from the Perseus galaxy cluster. In a recent study by Simionescu et al., Suzaku data were used in conjunction with the \texttt{\texttt{CLMASS}} and \texttt{\texttt{NFWMASS}} models. The results from these fits reveal that the baryon gas fraction of the outer regions of the cluster (exceeding 2/3 of the virial radius) exceeds the cosmic average baryon gas fraction \cite{simionescu_baryons_2011}. Such a result conflicts with accepted theory and provided additional motivation for this study of the Perseus Cluster. \\ 
\indent{}The archival data obtained for this study is from the XMM-Newton telescope. This telescope, in comparison to Suzaku, offers significantly better angular resolution due to XMM-Newton's better optics. Suzaku had two main benefits, a shorter focal length and a lower radiation environment. These advantages give Suzaku a lower background emission profile. Suzaku also has a smaller light collecting area compared to XMM. With a higher angular resolution, evidence of clumping will become significantly more apparent when applying mass determination and density models to derived spectra. The larger effective light collecting area also provides the benefit of higher photon counts over a shorter exposure length, providing better overall statistics. This should offer the capability to detect any clumping of matter in regions noted to have higher-than-anticipated baryon gas fraction.  If such clumping effects are observed, this could easily explain the conflict present in the Simionescu et al. study. Through this study, it is the goal of the investigator to supply ample reason for the continued study of diffuse and extended galaxy clusters using the analysis procedures to follow. 
\subsection{Expected Issues}	
\paragraph{}The study of Perseus using the XMM-Newton observatory poses several unique problems. The field of view of the instrumentation on board the telescope is roughly $30'$. The cluster itself has an extent of nearly 2$^{\circ}$, much larger than the FOV of the telescope. Because of the extent of the cluster, there is no region that can easily be selected to adequately model the background emission of the cluster. 
This issue is dealt with through the use of the ESAS data reduction package. This recently released analysis package provides several new tools to deal with background effects from extended diffuse X-Ray sources. This involves multiple tools designed particularly for the purpose of cleaning processed event files of soft proton, quiescent particle background, solar wind background, and crosstalk contamination (due to bright X-Ray sources located off-axis from the central observation heading). Furthermore, the novel use of ROSAT All-Sky Survey data as a spectral model to diffuse X-Ray background provides a powerful method to eliminate background emissions. 
Due to the very recent release of both the \texttt{CLMASS} and ESAS packages, this method of analysis has yet to be attempted on the Perseus galaxy cluster.

\section{Methodology}

\subsection{Initial Data Reduction}\label{initial}

The data processing techiques used to create spectra and images of the cluster
are based off of the XMM-Newton Extended Source Analysis Software (ESAS)
{ \cite{snowden_cookbook_2006}}. Significant improvements were made to the
workflow of this software distribution. In particular, a fully automated
framework was designed, based on work done in
{ \cite{wampler-doty_survey_2012}}, to streamline data processing. Mechanisms
to catch and eliminate errors were also introduced. This is
discussed in more detail in {\S}\ref{Automation}.

The ESAS software package first examines and produces a set of unprocessed and
uncleaned images directly from the detector with the \tmtexttt{odfingest}
routine. This is followed by the production of calibration files by the
\tmtexttt{cifbuild} routine, which allows reference calibration detector files
to be used in the production of cluster and background spectra as well as
image files.

\begin{figure}[h]
\begin{center}
 \includegraphics[width=.7\textwidth]{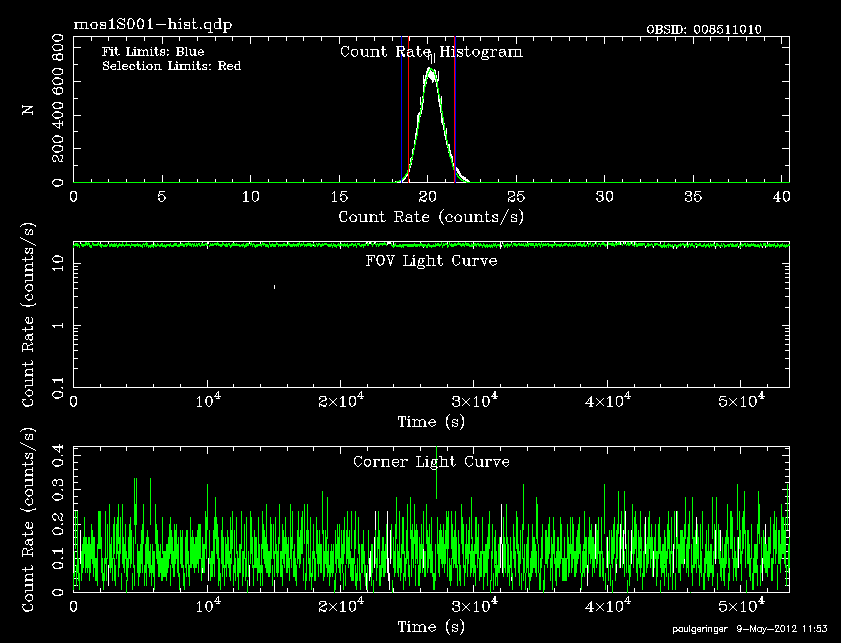}
  \caption{\label{light-curves}Light curves of the MOS1 detector for the first exposure of ObsID 0085110101.}
\end{center}
\end{figure}

A basic preparation task, \tmtexttt{emchain}, first generates the event
lists. Then, \tmtexttt{mosfilter} (another utility), examines the
unprocessed files for signs of obvious soft-proton contamination. It also
provides a master clean file that is used to derive spectra. The soft-proton
contamination characteristics can be examined directly through the plotting of
the observation histogram and light curves (see Fig. \ref{light-curves}).
Observations that contain low levels of soft-proton events tend to have a
light curve distribution that is generally uniform. Observations that do
experience significant soft-proton contamination tend to observe low levels of
high intensity events. Such contaminated observations in this study were
noted, as further spectra fitting would indicate whether the not reliable
results could be produced, or if the observation under study had to be
discarded.
\subsection{Masking and CCD Examination}\label{Masking}
\begin{figure}[h]
\begin{center}
 \includegraphics[width=.7\textwidth]{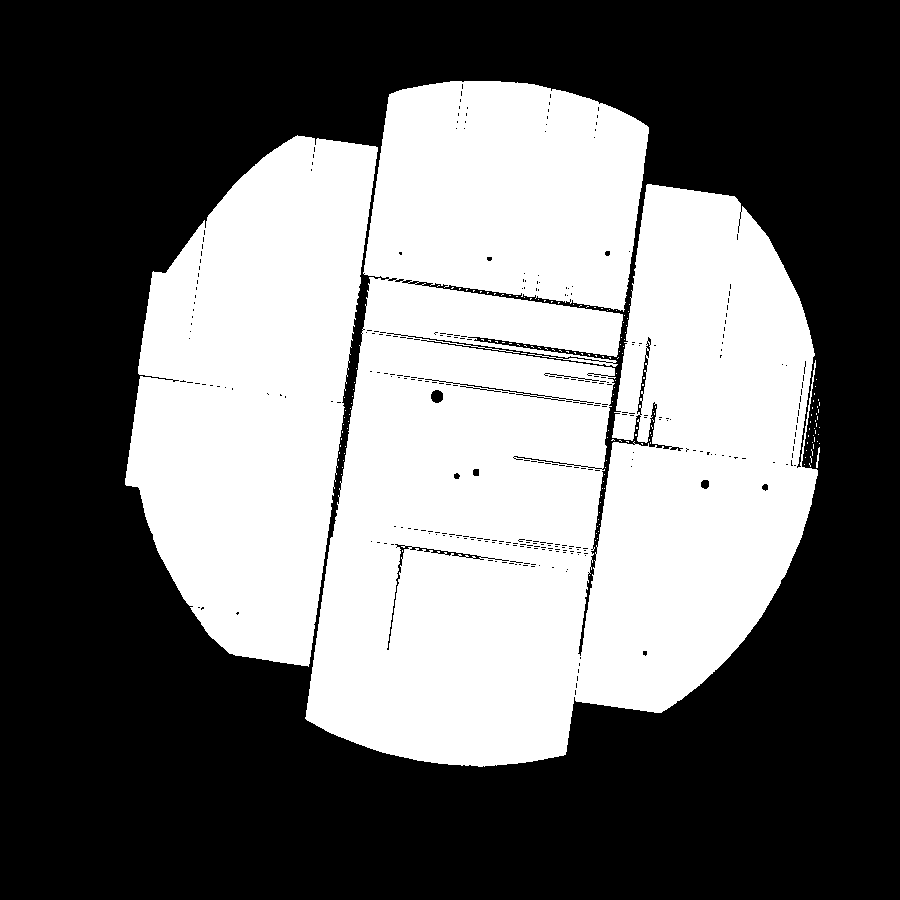}
  \caption{\label{cheese}Cheese mask image of ObsID 0085110101. This image indicates the overall profile of the MOS detectors.}
\end{center}
\end{figure}

The first step to producing spectra involved the creation of point source
mapping. As the primary processing techniques emphasized in ESAS are for diffuse
spectra, point sources can cause a significant level of contamination in the
resulting spectra. The ESAS \tmtexttt{cheese} task was used to exclude such
sources. The \tmtexttt{cheese} utility runs source-detection algorithms to
locate point sources, which are then added to a point source list. From this
list, image masks (see Fig. \ref{cheese}) are created to exclude these point
sources from final images.

Point sources were defined by the parameter selection of \tmtexttt{cheese}.
The threshold intensity value of an $10^{- 14}$ ergs/s cm$^2$ was used for the detection of potential point sources. Detected sources were then excised by
removing the region radially surrounding the source to half the
surface brightness of the background. The source detection algorithm was
limited so that sources within $40''$ of each other were not excluded, so as
to prevent brighter, extended sources from excision. However, this required the manual examination of the image for bright, closely-packed foreground sources, which the \texttt{cheese} task would skip. Any such sources were then excluded. 

For the purpose of this study, images were generated by \tmtexttt{cheese} and
all subsequent tasks. Masks generated by \tmtexttt{cheese} were made for the
range of energies between 400 eV through 10 keV.

Detector images of the observation were then examined for CCDs operating in
 anomalous states. CCDs behaving in this format tend to over-estimate low
energy ($<$ 1 keV) background. As the Perseus cluster has significant low
energy emission ($\sim$ 0.3 keV), exclusion of anomalous CCDs states was vital. CCDs were
determined to be in  anomalous states through the output of the corner hardness
ratios in the \tmtexttt{mos-filter} task as well as examining the images
visually. A shell script was used to extract the notably bad CCDs from the
output of \tmtexttt{mos-filter} and automatically set future tasks to
exclude them from processing.

\subsection{Annulus Generation}

\begin{figure}[h]
\begin{center}
\includegraphics[width=.45\textwidth]{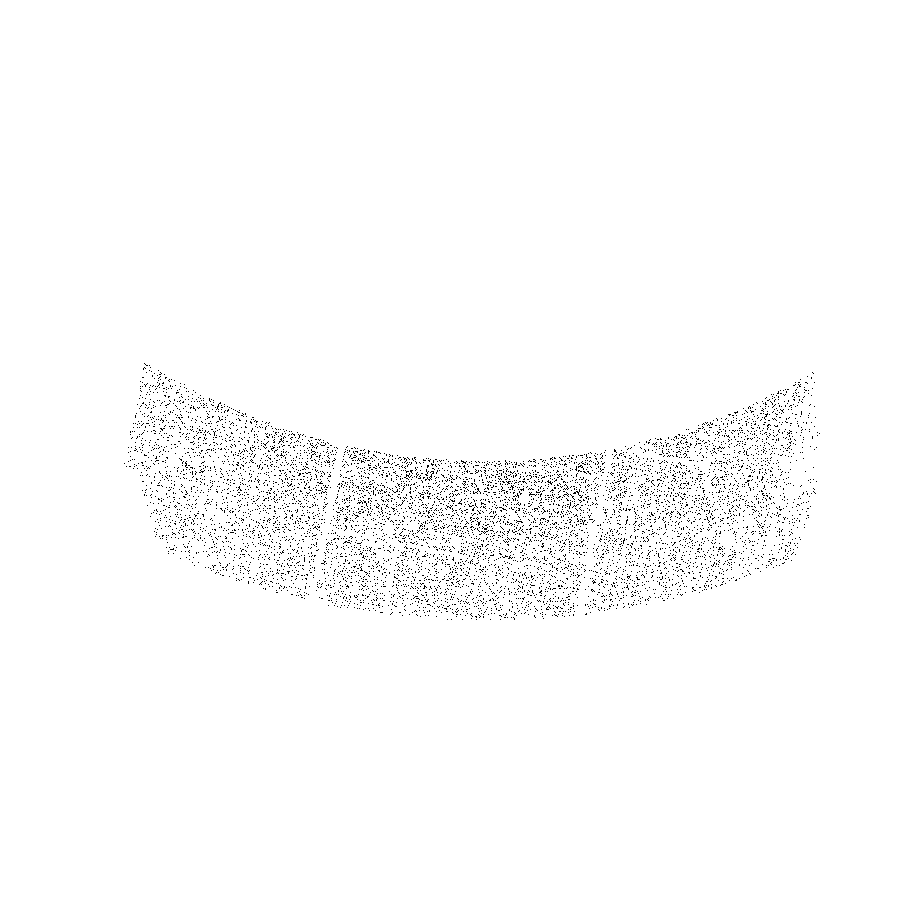} \includegraphics[width=.45\textwidth]{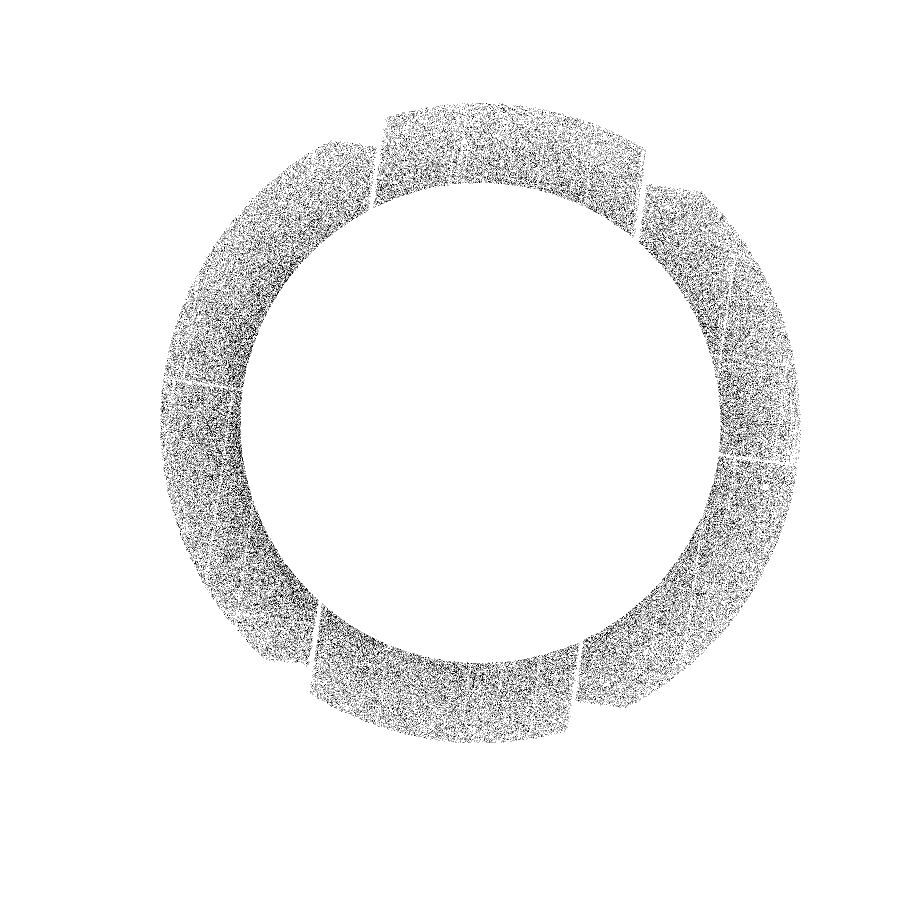}
  \caption{\label{annuli-part}Left: A partial annulus from a non-center observation. Right: A full annulus from the center cluster observation.}
\end{center}
\end{figure}

Production of the annular shaped sections was necessary for \tmtexttt{\texttt{CLMASS}}
and\tmtexttt{ \texttt{NFWMASS}} mixing model analysis. This required the creation of
several Python-based programs to create spectra across identical sky
coordinates for both MOS detectors. Instead of creating masking regions for
the included \tmtexttt{mos-spectra} and \tmtexttt{mos\_back} tasks, partial
annuli of events were computed directly from event files. This involved use
of the \tmtexttt{kapteyn} Python module and a novel rotation algorithm
described below. An example of the resulting event files can be seen in Fig.
\ref{annuli-part}. When generating an annular section, the events contained by
the annulus section were rewritten into a new event file containing the
identical FITS formatted header of the original event file.

\begin{figure}[h]
\begin{center}
\hspace{6em}\includegraphics[width=.8\textwidth]{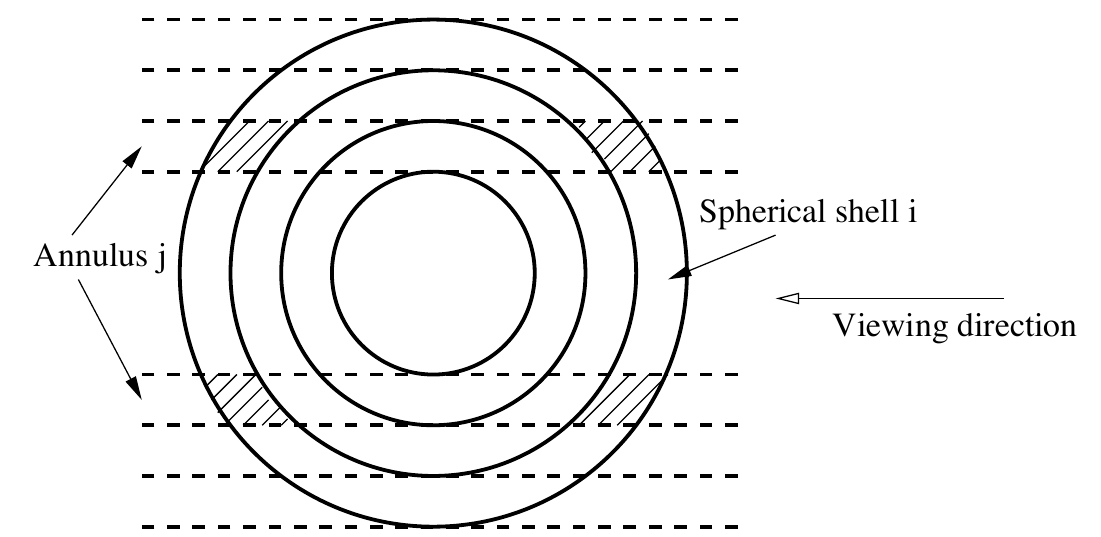}
 \caption{\label{shells} Diagram of spherical shells of cluster from \cite{nulsen_model-independent_2010}.}
\end{center}
\end{figure}

This program can be run in two modes, generating events of full cluster annuli
or annuli sections propagating in a specific radial direction from the cluster center. This system
permits a wide variety of potential analysis and lends itself particularly
well to the use of spectra in mass mixing models (such as \tmtexttt{\texttt{CLMASS}}
and \tmtexttt{\texttt{NFWMASS}}), which determine the characteristics of the spherical
shells defined by the cylinder of three-dimensional emission described by the
annuli (Fig. \ref{shells}). 

Event files that were generated were sorted according to the radial
spacing between the great arcs describing the annulus as well as the angle in
radians described by the flat edges of the annulus. This is particularly
useful when generating spectra of the full annulus encompassing the cluster,
as it determines the exact angular extent of each annulus. In the case of
directional analysis, this is particularly beneficial for any situation where
a direction chosen might lie along the edge of the observations field of view;
the angle is defined by the maximum and minimum right ascension of all events
within the reformatted event file. This angle generation is vital for the
use of many mixing models, as it is necessary to describe the particular region over which
to apply the model.

\subsubsection{Annulus Section Selection Algorithms}\label{annulusalgo}

Two annulus selection algorithms were used in this research:
\begin{itemizedot}
  \item \tmtextit{partialAnnulus}: Used to compute an annulus section $\mathcal{S}$ given an inner radius $r_1$, an
  outer radius $r_2$, and two bounding angles $\theta_1$ and $\theta_2$.
  
  \item \tmtextit{fullAnnulus}: Used to compute an annulus $\mathcal{A}$ given by radii $r_1$ and $r_2$, along with an associated bounding angles in the case that the annulus was not complete
  (i.e. when the annulus was generated from a peripheral, non-center
  observation).
\end{itemizedot}

\begin{table}[!h]
\begin{center}
  \begin{tabular}{l}
    \tmtextit{partialAnnulus}($evtfile, \rho_0, \delta_0, r_1, r_2, \theta_1, \theta_2$):\\
    {\hspace{3em}}\emph{rotevts} $\leftarrow$ \textbf{map} $rot(\rho_0, \delta_0, \cdot, \cdot)$ \textbf{on} $evtfile$  \\
    {\hspace{3em}}\textbf{return} $\{ \rho', \delta' \in rotevts : \theta_1 \leq \rho' \leq \theta_2 \ \& \  r_1 \leq \delta' \leq r_2 \}$ 
 %   {\hspace{3em}}\textbf{return} $annulus$  \\
  \end{tabular}
 \end{center}
  \caption{\label{pseudo-annulus-part}A \emph{pseudo-code} implementation of the
  \tmtextit{partialAnnuli} routine in mathematical shorthand.}
\end{table}

\begin{table}[!h]
\begin{center}
  \begin{tabular}{l}
    \tmtextit{fullAnnulus}($evtfile, \rho_0, \delta_0, r_1, r_2$):\\
    {\hspace{3em}}\emph{rotevts} $\leftarrow$ \textbf{map} $rot(\rho_0, \delta_0, \cdot, \cdot)$ \textbf{on} $evtfile$  \\
    {\hspace{3em}} $ \mathcal{A} \leftarrow \{ \rho', \delta' \in rotevts : r_1 \leq \delta' \leq r_2 \}$ \\
     {\hspace{3em}}$\displaystyle\theta \leftarrow \max_{\rho', \delta' \in \mathcal{A}} \rho' - \min_{\rho', \delta' \in \mathcal{A}} \rho' $\\
    {\hspace{3em}}\textbf{if} $\theta > \pi :$\\
    {\hspace{6em}}$\displaystyle\theta \leftarrow \max_{\rho', \delta' \in \mathcal{A}} (\rho' + \pi\ \mathbf{mod}\  2\pi) - \min_{\rho', \delta' \in \mathcal{A}} (\rho' + \pi\ \mathbf{mod}\  2\pi) $ \\
    {\hspace{3em}}\textbf{return} $\mathcal{A}, \theta$  
  \end{tabular}
 \end{center}
  \caption{\label{pseudo-annulus-full}A \emph{pseudo-code} implementation of the
  \tmtextit{fullAnnuli} routine in mathematical shorthand.}
\end{table}

\paragraph{} These two pieces of \emph{pseudo-code}, programmed in Python, provide the ability to make precise event regions of multiple radial distances simultaneously. Partial annuli are particularly useful for directional analysis, or for selecting very narrow regions for analysis without the use of masking files. The full annuli code is particularly useful in central cluster observations, where the creation of single complete annulus is possible. An annulus defining system is necessary for the use of cluster mass mixing models. 

The rotating algorithm may be described as follows: Let $\rho_0,
\delta_0$ be the right ascension and declination, respectively, of the center
observation. The position of these coordinates on the unit sphere is the
vector $z' \equiv \langle \cos \rho_0 \cos \delta_0, \sin \rho_0 \cos \delta_0
\nocomma \nocomma \nocomma, \sin \delta_0 \rangle$. As we want this
coordinate to be rotated to the north pole of the unit sphere, it is desired
to rotate the point around a pivot on the equator of the unit sphere. The
coordinates of this pivot point are $\rho_0 + \pi / 2$ right ascension, 0
declination. The vector coordinates on the unit sphere of the pivot point are
$y' \equiv \langle - \sin \rho_0, \cos \rho_0, 0 \rangle$. The crossproduct is
given by:
\begin{eqnarray*}
  x' & \equiv & z' \times y'\\
  & \equiv & \langle - \cos \rho_0 \sin \delta_0, - \sin \rho_0 \sin
  \delta_0, \cos \delta_0 \rangle
\end{eqnarray*}
Together $x'$, $y'$ and $z'$ form a 3 dimensional rotation matrix $R$
\begin{eqnarray*}
  R & = & \left(\begin{array}{c}
    x'\\
    y'\\
    z'
  \end{array}\right)\\
  & = & \left(\begin{array}{ccc}
    - \cos \rho_0 \sin \delta_0 & - \sin \rho_0 \sin \delta_0 & \cos
    \delta_0\\
    - \sin \rho_0 & \cos \rho_0 & 0\\
    \cos \rho_0 \cos \delta_0 & \sin \rho_0 \cos \delta_0 & \sin \delta_0
  \end{array}\right)
\end{eqnarray*}
The action of this matrix rotates $\rho_0, \delta_0$ to the $z$ axis and
$\rho_0 + \pi / 2, 0$ to the $y$ axis. To rotate a particular event with
coordinates $\rho \nocomma, \delta \nocomma,$and corresponding unit vector
$\hat{v}$, the dot product $R$:
\begin{eqnarray*}
  R \cdot \hat{v} & = & \left(  \begin{array}{c}
    \sin \delta \cos \delta_0 - \sin \delta_0 \cos \delta \cos \left( \rho -
    \rho_0 \right) \\
    \cos \delta \sin \left( \rho - \rho_0 \right) \\
    \cos \delta \cos \delta_0 \cos \left( \rho - \rho_0 \right) + \sin \delta
    \sin \delta_0
  \end{array}  \right)
\end{eqnarray*}
It can be seen that $\rho'$, the rotated right ascension of the event being
rotated up to the north pole is given by:

\begin{eqnarray*}
  \rho' & = & \arctan \left( \frac{\cos \delta \sin \left( \rho - \rho_0
  \right) }{\sin \delta \cos \delta_0 - \sin \delta_0 \cos \delta \cos \left(
  \rho - \rho_0 \right) } \right)
\end{eqnarray*}

In the interest of computing annuli sections, it is necessary to compute the
great circle distance of $\rho, \delta$ from $\rho_0, \delta_0$. This is given
by the {\tmem{Vincenty great circle distance formula}} \cite{vincenty_direct_1975} which is associated
with the new declination of the rotated event:
\begin{eqnarray*}
  \delta' & = & \arctan \left( \frac{\sqrt{\left( \cos \delta_0 \sin \left(
  \rho - \rho_0 \right) \right)^2 + \left( \cos \delta \sin \delta_0 - \sin
  \delta \cos \delta_0 \cos \left( \rho - \rho_0 \right) \right)^2}}{\sin
  \delta \sin \delta_0 + \cos \delta \cos \delta_0 \cos \left( \rho -
  \rho_0 \right)} \right)
\end{eqnarray*}
Since both $\rho'$ and $\delta'$ have many terms in common, they can be
computed together efficient.

The resulting algorithm is implemented in the \tmtexttt{rot} function in the
\tmtexttt{filter\_annuli.py} module. It is describe below using
\tmtextit{pseudo-code} in Table \ref{pseudo-rot}.

\begin{table}[h]
\begin{center}
  \begin{tabular}{l}
    \tmtextit{rot}($\rho_0, \delta_0, \rho, \delta$):\\
    {\hspace{3em}}$\rho' \leftarrow \arctan \left( \frac{\cos \delta \sin
    \left( \rho - \rho_0 \right) }{\sin \delta \cos \delta_0 - \sin \delta_0
    \cos \delta \cos \left( \rho - \rho_0 \right) } \right)$\\
    {\hspace{3em}}$\delta' \leftarrow \arctan \left( \frac{\sqrt{\left( \cos
    \delta_0 \sin \left( \rho - \rho_0 \right) \right)^2 + \left( \cos
    \delta \sin \delta_0 - \sin \delta \cos \delta_0 \cos \left( \rho -
    \rho_0 \right) \right)^2}}{\sin \delta \sin \delta_0 + \cos \delta
    \cos \delta_0 \cos \left( \rho - \rho_0 \right)} \right)$\\
    {\hspace{3em}}\tmtextbf{return} $\rho', \delta'$
  \end{tabular}
  \end{center}
  \caption{\label{pseudo-rot}A \emph{pseudo-code} implementation of the
  \tmtexttt{rot} Python function}
\end{table}

\subsection{Objective and Background Spectra Generation}\label{spectragen}

Spectra were generated using the ESAS mos-spectra task. CCDs to be excluded
were automatically loaded as described in {\S}2.2. Instrument and exposure
number were also automatically generated from processed data. For all
observations of Perseus, images were generated along with the spectra for the
energy ranges of 400 eV through 1.2 keV and 2.0 keV through 7.2 keV. Such image
generation is optional and time intensive, but for the purposes of this paper,
image generation was necessary to provide redundancy and confirmation than
correctly formatted annuli sections were generated. The \texttt{mos-spectra} task
uses the now rotated and redefined event files to produce the annular-shaped spectral regions as
well as necessary intermediate files for the second task, \texttt{mos-back}, which
generates the model particle background spectra.

Ancillary response files (ARF) and response matrix files (RMF) are also
created for each spectra generated. The ARF files provide the ability to use
the spectra generated in any standard high-energy spectra analysis software by
calculating the effective area of the detector in question, while the RMF
similarly follows preset standard conventions to describe the actual behavior
of the detector as a function of energy, through the use of the calibration
files (\S\ref{initial}).

Object spectra, background spectra, ARF, and RMF files are grouped together as
one main spectra file through the use of the \texttt{grppha} program. This permits the
quick and easy loading of all the necessary spectral files into whichever
analysis program is to be used, simply by specifying the grouped spectra file.

\subsection{Automation Framework}\label{Automation}

The tasks listed in the previous sections consume a large portion of time and computer power to
properly generate. However, the process itself can run into errors in various
places. The use of the GNU \texttt{make} automation program was indispensable. It
permitted this investigator to create a single recipe outlining each step that
needed to be taken before the next task in the pipeline should be run. At the
end of each task, a file would be generated should no fatal errors occur. The
existence of these "completion files" were listed as dependencies on the next set of processing
tasks. If the file wasn't created, the program would halt analysis at the point
of failure. This permitted extensive debugging procedures that previously
weren't available, as well as greatly reducing user error while compiling the
parameters necessary for each task. This modular design is
extremely portable and can be run on nearly any UNIX-based computer, greatly
simplifying the ESAS procedure. One major benefit of this system is the
capability to do batch processing for numerous spectra at once, which reduced
processing time significantly.

\subsection{Spectra Analysis}

\subsubsection{Spectra Preparation}

Spectra were analyzed using the 12.7.0 release of the \texttt{XSPEC} spectral analysis
software package. Local \texttt{\texttt{NFWMASS}} and \texttt{\texttt{CLMASS}} models were used for the purposes
of this analysis. Cosmological constants in use were $H_0 = 70$, $\Omega_{\Lambda} =
0.73$, $\Omega_m = 0.27 $, and relative solar abundance used for all abundance fitting parameters. Systematic
error was fixed at 0.

Spectra were loaded in groups of three or four as defined by the observation
the spectra originated from. The spectra analyzed were derived from the full
annulus mode of the spectra generation tasks (\S\ref{spectragen}), but were grouped by
observation to preserve directionality.

Bad event responses were ignored immediately. Any events below the range of
.6 keV and above 8.0 keV were excluded from fitting, due to the falloff at
higher energies and the non-standard response of the MOS detectors below $\sim 0.4$ keV. \

\subsubsection{Fitting Procedures}

Fits were run using the standard statistical package loaded into \texttt{XSPEC};
$\chi^2$ statistics were used for the fitting.

All fits were done after first varying to parameter to be fit until the
resultant $\chi^2$ value was reasonably decreased from the default parameter
value. In cases where background parameters were fit, the program was left to
attempt a fit without any adjustment initially, due to the large number of
parameters needed to be varied simultaneously to achieve a good fit ($\sim 40$
parameters).

\subsubsection{Soft Proton, Cosmic Background, and Instrumental effects}

As described in the ESAS Cookbook \cite{snowden_cookbook_2006}, the background was modeled by
multiple components for local foreground, instrumentation, and cosmic
background sources. Gaussian model components were used to describe the two
radiation sources (Al K-$\alpha$ and Si K-$\alpha$), with energies of roughly
1.49 and 1.75 keV. For each spectra, a power-law model was added with a photon
index fixed at 0.2 across all spectra, as recommended. These
power-law components were not folded into the instrument effective area,
depending only upon the RMF file of each spectra.

Cosmic background was modeled through the use of ROSAT All-Sky Survey
archival data. An annulus of the region between 1 and 1.5 degrees from the
center of the cluster was added to the loaded spectra and fit directly for the
first four channels of the detector \cite{snowden_cookbook_2006}. This provided the necessary low energy
cosmic background filtering. The models applied to this spectra included a low
energy ($<$ .1 keV) thermal \texttt{APEC} component (unabsorbed), an absorbed thermal
\texttt{APEC} component of a slightly higher energy ($\sim$.25 keV), and an absorbed
power law fixed at a photon index of 1.46 to account for background
cosmological X-Ray sources (normalization fixed to the canonical value
$8.88\times10^{7}$). Once the background model parameters were well fit, the RASS
spectrum itself was excluded, as the spectra used interfered with the fitting
of the mixing models used in the mass analysis.

Two Gaussian components, centered at 0.65 keV and 0.56 keV were also included to
account for solar wind-ionized particle contamination, but almost universally
these models return extremely low, non-fitting normalizations and were locked at zero
intensity.

\subsubsection{\texttt{APEC} and \texttt{VAPEC} Cluster Emission fitting}\label{clusteremission}

Once the background had been well fit, an absorbed thermal \texttt{APEC} component was
loaded as the primary model for all source spectra. Redshift for these \texttt{APEC}
components was fixed at the standard valued for Perseus of 0.0183  \cite{sanders_deeper_2007,fabian_deep_2003}.
Abundances and temperatures for each spectra were allowed to vary freely. The absorbed component was set to values
derived from the X-Ray Background Tool used to generate the ROSAT background
spectrum.

Once the standard \texttt{APEC} model was reasonably well fit, with a reduced $\chi^2$ of $\sim$ 1, it was replaced with a
\texttt{VAPEC} model. A \texttt{VAPEC} model allows more precise control of the abundances of 12
different elements. For the purpose of this study, all alpha elements \cite{matteucci_chemical_1995}
were linked to the abundance of oxygen, which was allowed to vary, while Fe was
allowed to vary freely. All other elemental abundance parameters in the \texttt{VAPEC}
model were frozen at zero. Similar to the \texttt{APEC} model, all normalizations were
frozen to the innermost annuli spectra.

\texttt{APEC} models were chosen due to the ability to select one central abundance as well as the ability to switch between the standard model and the variable abundance model with ease. Further, the \texttt{APEC} model is based upon the principle that the source emission, which is an ionized plasma contained in the ICM of Perseus, was ionized through collision processes. Such a model, based upon thermal bremsstrahlung and line emission processes, is a reasonable emission model for the Perseus cluster \cite{arnaud_temperature_1994}.

\subsubsection{Mixing Model Fitting}

Following the fitting of the \texttt{APEC} and \texttt{VAPEC} models, the models were further
edited to include the mixing models \texttt{NFWMASS} \cite{navarro_universal_1996}
and the \texttt{CLMASS} model. 

\texttt{NFWMASS} was initially fit as a simple test to confirm
the spectra were properly formatted. In one case, the spectra had some slight
overlap between annuli sections, resulting in errors. This spectra was excluded from all analysis. To provide a good fit
for the mass density $\rho_0/(\frac{r}{a} (1 + \frac{r}{a})^2)$ and total mass of the region $M_{total}$ (Equations \eqref{nfwpot} and \eqref{totalmass}), temperatures were linked to the proper temperature components of the
model, while the scale length $a$ and potential $U$ were varied manually initially
to roughly expected values, to ensure that the fitting procedure would
converge. The thermal models were absorbed \texttt{APEC} models (\S\ref{clusteremission}), as
opposed to \texttt{VAPEC} models, so as to reduce the uncertainty of the large number
of parameters in the \texttt{VAPEC} model.
\begin{eqnarray}
\label{nfwpot}
U &=& 4\pi G \rho_0 a^2 \mu m_H\\
\label{totalmass}
M_{total} &=& \int \frac{4\pi r^2 \rho_0 \  dr}{\frac{r}{a} (1 + \frac{r}{a})^2}
\end{eqnarray}

Once \texttt{NFWMASS} models were well fit, the \texttt{NFWMASS} component was exchanged for a
\texttt{CLMASS} component, for the derivation of electron number density and gas
densities. Similar temperature linking procedures were utilized, while
densities for each shell were allowed to vary freely after being set to
roughly expected values \cite{simionescu_baryons_2011}. Densities for each region were produced in the units of (\ref{clmassden}):
\begin{equation}
\label{clmassden}
\rho = \frac{keV} {G \mu m_H u^2}
\end{equation}
where $u$ is in units of arcseconds. Conversions to cgs units can be found in \S\ref{calculations}.

In each case, beta models were not used, as the outer annulus was taken to be
the approximately outer emission from the cluster. Normalizations for each spectra were tied to the normalization of the
innermost annuli, as this normalization describes the flux through all of the
shells of the spherical cluster. The inner radial distance
from the center of the cluster were in units of arcseconds. To describe the
angular extent of each region the \textbf{AREASCAL} keyword within the spectra file was
edited to the appropriate value for each spectra, while the outer edge radial
distance of each annuli was stored in the \textbf{XFLT001} keyword, as specified \cite{nulsen_model-independent_2010}.

\section{Results}

\subsection{Tabulating Results}
\indent{}From Table \ref{nfwclmasstab}, it can be seen that much of the analysis was plagued by high levels of uncertainty in the parameter fit, despite generally acceptable $\chi^2$ statistics. In the case of the \texttt{VAPEC} model (Table \ref{vapectab}), this was generally not a significant issue, though it was apparent that the linking of so many alpha elements together added an unnecessary level of uncertainty into that parameter. Iron abundances tended to be reasonably well fit by comparison, with significantly lower uncertainties in almost all regions.
\texttt{CLMASS} and \texttt{NFWMASS} model fits experience high levels of error (Table \ref{nfwclmasstab}). In comparing density, gas fraction, and baryon fraction results with Simionescu et al. and Sanders et al., it is apparent that the results obtained were generally within a factor of ten of the previous studies' reported results (see \S\ref{calculations}). 
\begin{table}[!h]
\begin{center}
\includegraphics[width=\textwidth]{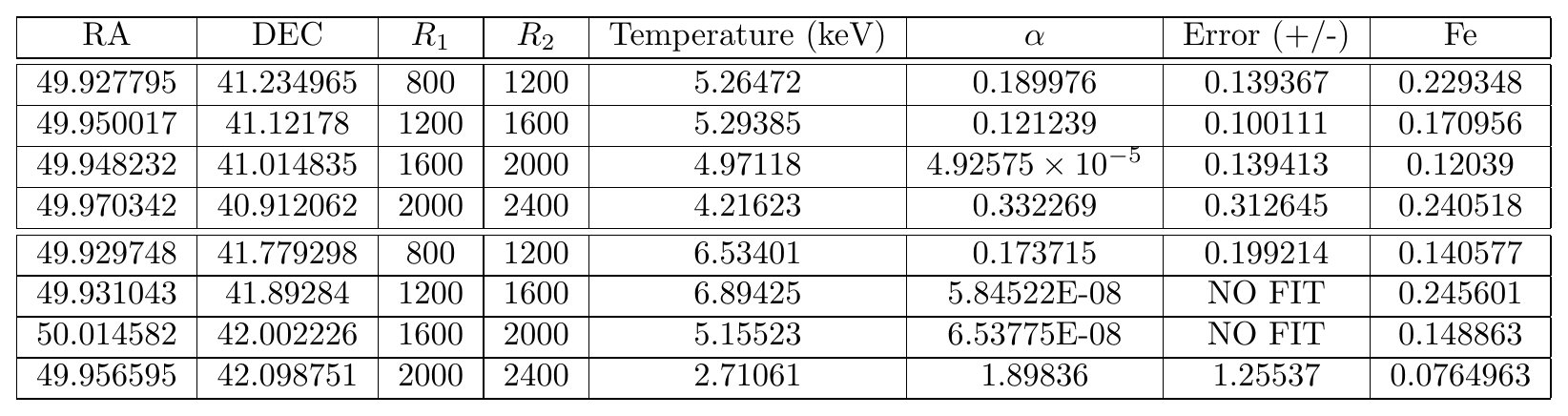}
  \caption{Sample of \texttt{VAPEC} model temperature, $\alpha$ and Fe  abundance fit from ObsID 0085590201.}\label{vapectab}
  \end{center}
\end{table}
\begin{table}[!h]
\begin{center}
\includegraphics[width=\textwidth]{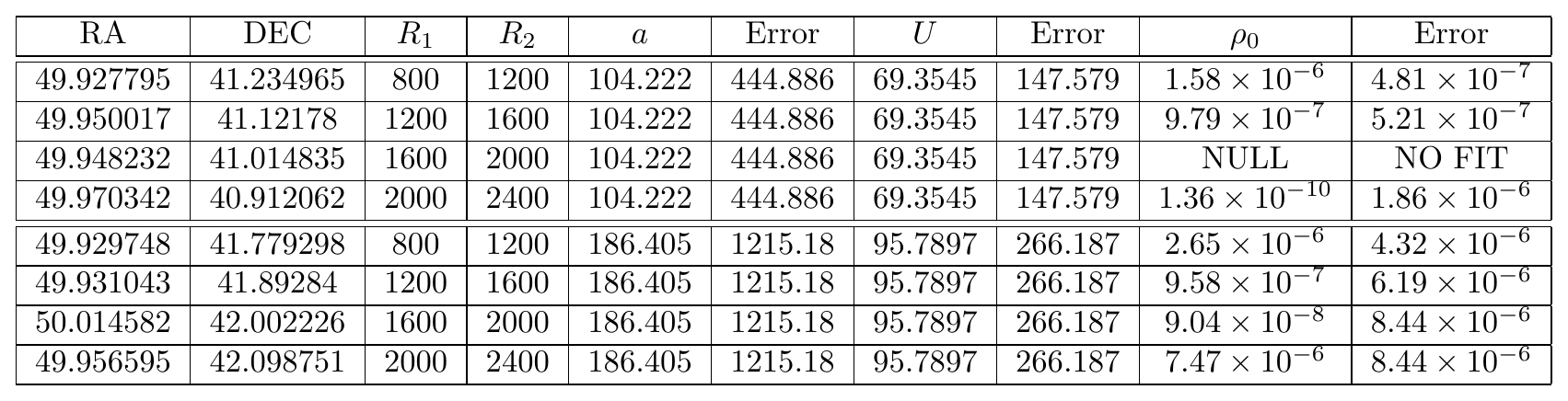}
 \caption{\texttt{NFWMASS} fit for $a$ and $U$ and \texttt{CLMASS} fit for $\rho_0$ from ObsID 0085590201 and ObsID 0305690301 respectively. $a$ and $U$ are the same for all annuli of each observation, as only one value is provided for each spectral group.}\label{nfwclmasstab}
  \end{center}
 \end{table}

\subsection{Calculation}\label{calculations}
\indent{}To generate the final density and mass numbers the annular regions, from the provided numbers generated by the \texttt{NFWMASS} and \texttt{CLMASS} models, a few calculations had to be made to first. Angular and luminosity distances were calculated to provide a value of roughly .367 kpc/arcsec and $3\times10^{21}$ meters
 \cite{wright_cosmology_2006}. This conversion factor was used both in the calculation of the gravitating mass density derived from \texttt{CLMASS} as well as gravitational potential from \texttt{NFWMASS}. A constant factor of $1.6\times10^{-9}$ in units of ergs/keV was included. The value of 0.62 is used for the mean mass fraction of a full ionized plasma \cite{allen_allens_2000}, while $1.6\times10^{-24}$ grams is used as the mass of a proton. \\

The gas density from \texttt{CLMASS}, in cgs units of grams/cm$^{3}$, is given by the equation:
\begin{eqnarray*}
\rho_g = \frac{1.6\times10^{-9}\cdot{\rho_0}}{G(.6\cdot1.6\times10^{-24})\cdot(.367\cdot3\times10^{21})^{2}}
\end{eqnarray*}

The electron number density is given by the equation:
\begin{eqnarray*}
n_e \equiv N_e / V = \frac{\rho_g}{1.6\times10^{-24}}\times(1 - \frac{m_e\cdot N_e}{m_H\cdot N_H}) \cdot \frac{N_e}{N_H} \approx \frac{\rho_g}{1.6\times10^{-24}}
\end{eqnarray*}

While the total mass density of the region is provided (via \texttt{NFWMASS}):
\begin{eqnarray*}
\rho_t = \frac{1.6\times10^{-9}\cdot{U}}{G(.6\cdot1.6\times10^{-24})\cdot(a\cdot.367\cdot3\times10^{21})^2}
\end{eqnarray*}

Finally, the baryon fraction is provided by:
\begin{eqnarray*}
f_{gas} \equiv \frac{\rho_g}{(\rho_t - \rho_g)}
\end{eqnarray*}

\begin{figure}[!h]
\begin{center}
\includegraphics[width=\textwidth]{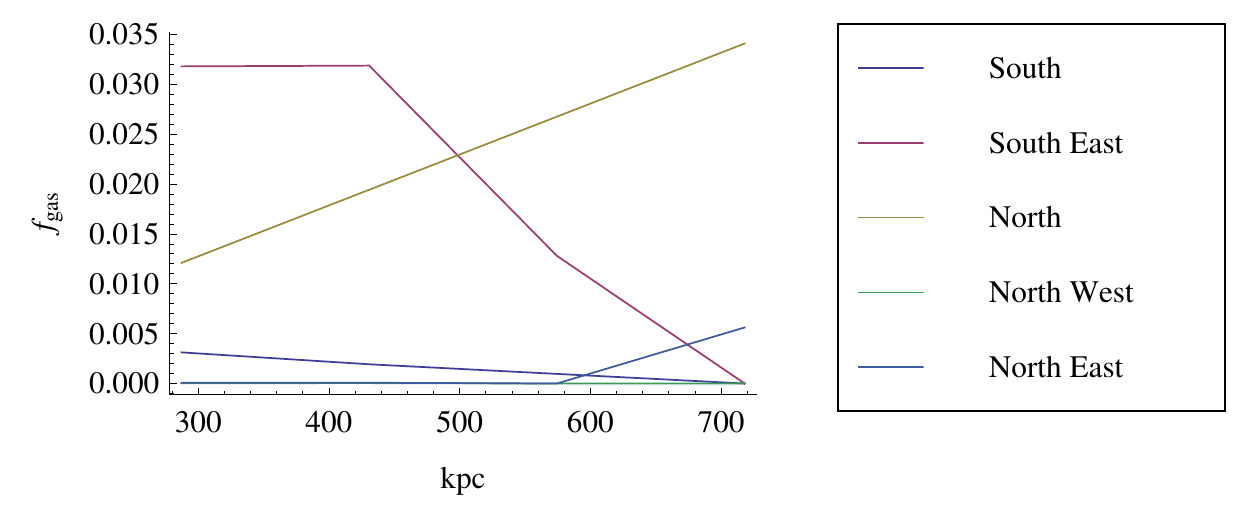}
\end{center}
\caption{\label{fgas} The dependence of the deprojected $f_{gas}$, the baryon gas fraction, as a function of radius. High levels of error skew this results an order of magnitude too low.}
\end{figure}

\indent{}In comparison to previous studies of the Perseus cluster, notably the aforementioned study by Simionescu and a related study by Sanders et al.  \cite{simionescu_baryons_2011,sanders_deeper_2007}, it has been determined that the error from the anticipated results were roughly of an order of magnitude higher, in the case of gas mass density, and an order of magnitude too low for the gas fraction (Fig. \ref{fgas}). However, due to the expectations of the sources of error and the extent to which these errors could alter these results, these error margins are considered to be acceptable. More rigorous practices must be utilized in future study of this region.

\begin{figure}[!h]
\begin{center}
\includegraphics[width=\textwidth]{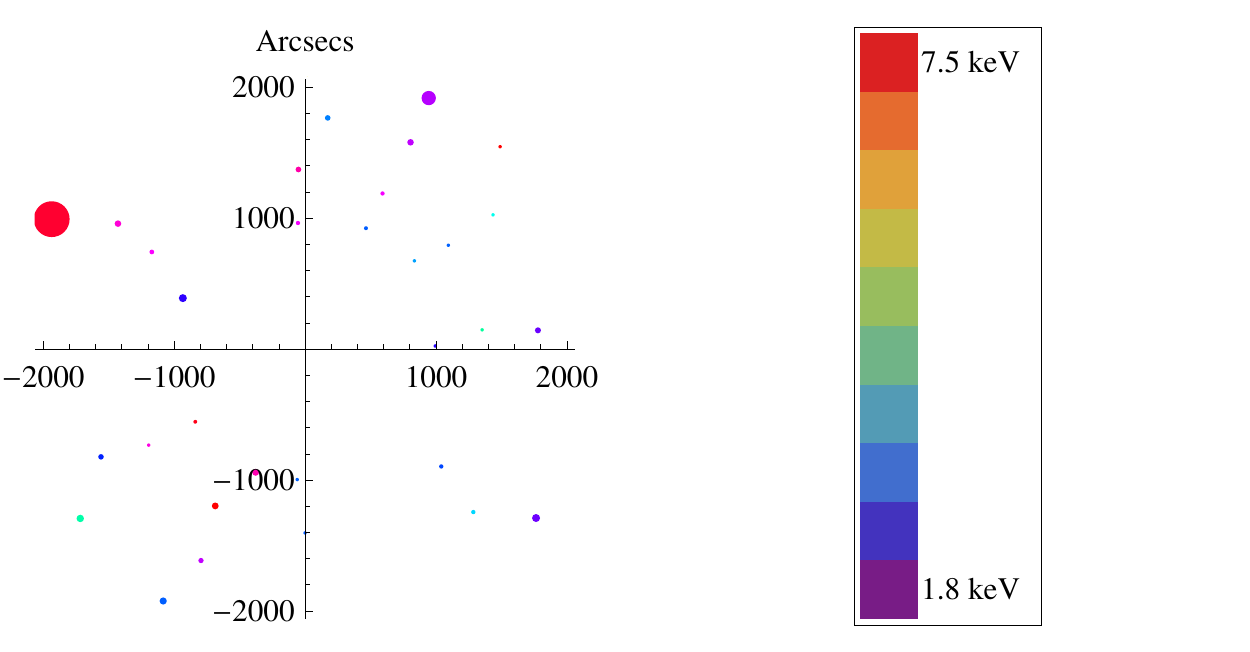}
\end{center}
\caption{\label{tempprofile} A profile of the projected temperatures across the Perseus cluster. Circle size is indicative of the level of error of the parameter value. Cluster center is at the origin. The code referenced in \S\ref{annulusalgo} was used to provide the coordinate rotation to produce both Fig. \ref{tempprofile} as well as Fig. \ref{abundprofile}. North is in the direction of the upper left quadrant. }
\end{figure}

\begin{figure}[!h]
\begin{center}
\includegraphics[width=\textwidth]{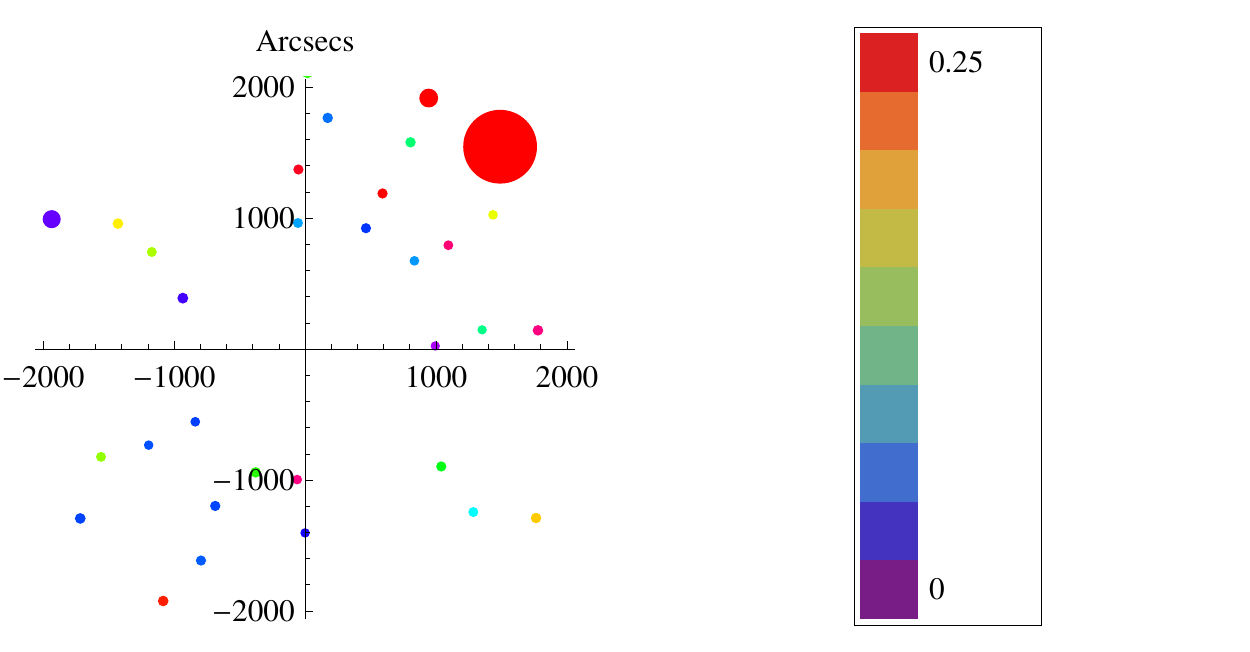}
\end{center}
\caption{\label{abundprofile} A profile of the Fe abundance across the Perseus cluster. Circle size is indicative of the level of error of the parameter value.}
\end{figure}

\indent{}In the case of temperature profiling and iron abundance, via the \texttt{VAPEC} model, results were far closer to presently accepted values to within a respectable margin of error \cite{arnaud_temperature_1994}. However, it is desirable to produce higher quality results for across the entire region, as the size of the annular sections used in this study are of significant breadth compared to the whole of the observed clustered. This provides a major limitation on the ability to pinpoint exact regions and define temperatures, as the temperatures and elemental abundances produced by the \texttt{VAPEC} model are not provided as a direct function of radius. Increasing the number of annular regions, as mentioned in \S\ref{future}, is expected to provide a continuous function describing the temperature and abundances dependent on direction and radius as well as significantly improved statistics.
%\begin{figure}[!h]
%\begin{center}
%\includegraphics[width=\textwidth]{massdensity.pdf}
%\end{center}
%\caption{\label{massdens} The dependence of $\rho_{g}$ as a function of radius.}
%\end{figure}
%\subsection{Arguments Against Hydrostatic Equilibrium}
One significant area of concern, as mentioned below in \ref{error}, is the distinctly non-uniform temperature distribution of the cluster (Fig. \ref{tempprofile}). This, tied with the spread of abundances observed (Fig. \ref{abundprofile}), provides reason to argue that the Perseus cluster does not model a system in hydrostatic equilibrium  \cite{fabian_deep_2003,fabian_distribution_1981}. However, it is expect that further study (\S\ref{future}) should yield results with more reasonable error to combat this potential failing of the mass mixing models. 

\indent{}As described in  \cite{sanders_deeper_2007}, non-spherically symmetric bubbles and ripples in the immediate central region of the cluster exist, while in some cases shock waves extending out to $\sim$170 kpc. This could particularly be indicative of recent interactions and collisions with galaxies or the infall of large quantities of matter into the central AGN. Furthermore, there is recent evidence of significant gas motion within the immediate 100 kpc of the central AGN  \cite{churazov_xmmnewton_2004}. With these considerations, it is possible that some error in the model fitting stems from the failure of the initial assumption concerning the equation of state of the cluster itself.

\subsection{Error Considerations}\label{error}
\indent{}While the error observed in this study was seen to be unusually high, much of the error, it is believed, can be reduced in future analysis. Only one detector was used in this analysis, though spectra were produced for both the MOS1 and MOS2 detectors initially. This current work on the MOS1 detector was done simply as a test case to ensure that results obtained from this new method were not significantly different, to the point where this method of analysis should be abandoned. The elimination of the second detector poses unique problems. Statistics were greatly reduced by only using half of the data, averaging results across the two detectors was not an option, and the inadequate performance of the primary detector due to physical damage. 

\begin{figure}[!h]
\begin{center}
\includegraphics[width=\textwidth]{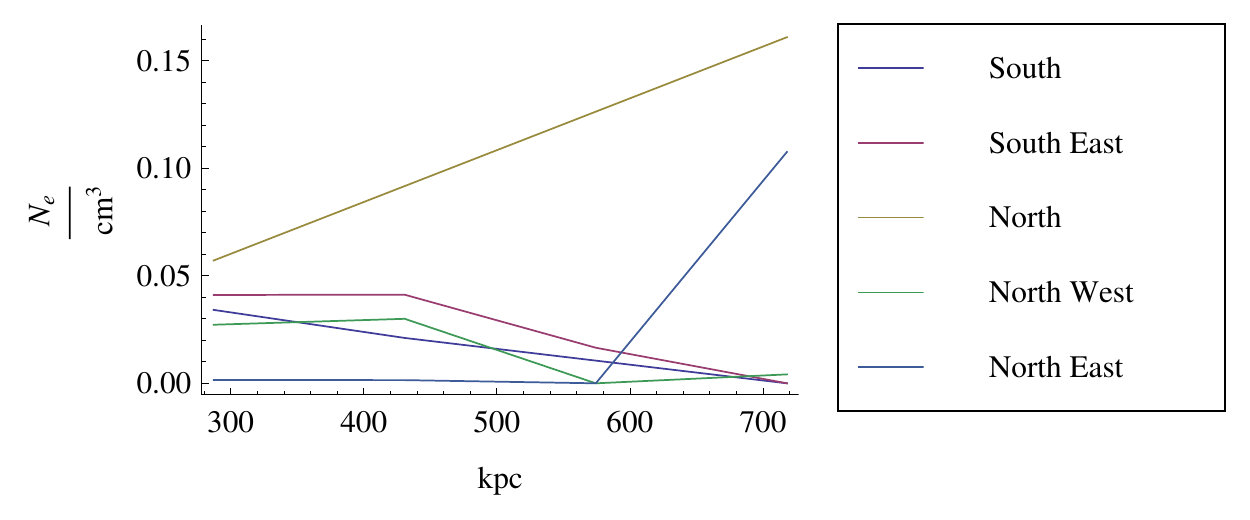}
\end{center}
\caption{\label{numbdens} A profile of the electron number density across the Perseus cluster. In combination with Fig. \ref{tempprofile} and Fig. \ref{abundprofile} are the main examples of high error stemming from the \texttt{NFWMASS} and \texttt{CLMASS} model fits.}
\end{figure}

\begin{figure}[!h]
\begin{center}
\includegraphics[width=.45\textwidth]{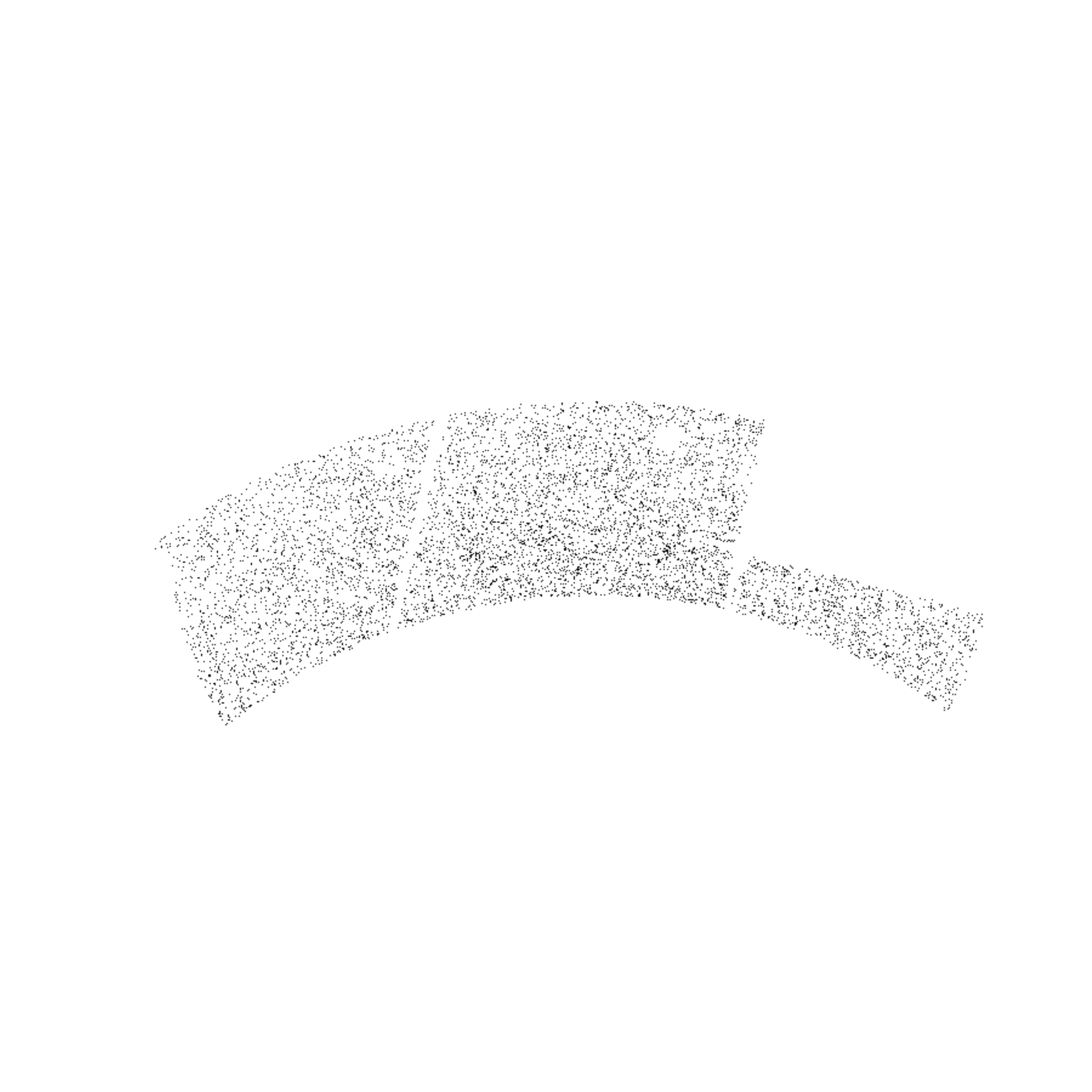} \includegraphics[width=.45\textwidth]{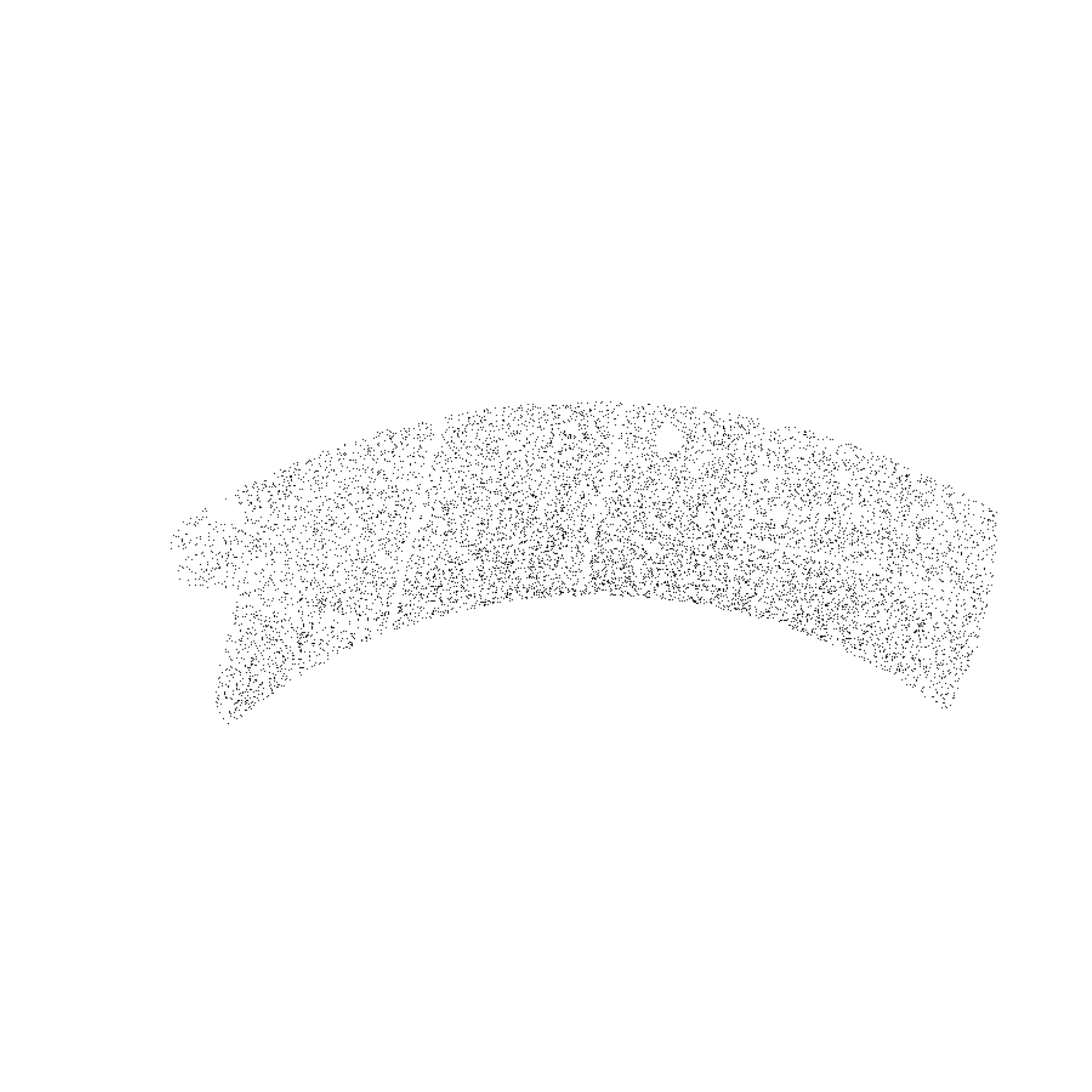}
\caption{\label{ccd_damage} On left: Image from the annuli across $1200''$ through $1600''$ with damaged CCD region in upper right corner. On right: Image from annuli across $1200''$ through $1600''$ for the intact MOS2 detector. Exposure retrieved from ObsID 0305690301.}
\end{center}
\end{figure}

\indent{}The detector used in this study was damaged in a micrometeorite impact, resulting in the total loss of the CCD6 of the MOS1 detector. In normal operation, such a loss would not be of significant detrimental effect to the quality of observation data; annular sections could be lacking significant amounts of the effect light collecting area (Fig \ref{ccd_damage}).  This is obvious when the statistical robustness of the Fe abundance and temperature fitting from the \texttt{VAPEC} models is considered, as this model doesn't rely upon the annular shape for fitting. Large gaps in the annular sections are expected to have been a major source of error for the \texttt{CLMASS} and \texttt{NFWMASS} models, as mixing models are dependent upon having a continuous set of annuli from a defined inner radius to the edge of the cluster (or to the farthest radial extend possible). 

\indent{}Another potential source of induced error was the chosen spacing and angular size of the annular regions. For the purpose of producing spectra that would be guaranteed to have adequate numbers of events, large regions, some annuli sections with $ > $ 30,000 events were used. The parameters for the annular sections were set to have radial extent of $400''$, while the angular size varied between $\frac{\pi}{6}$ to roughly $2\pi$ radians. In the course of the analysis, it was determined that having too few spectra for each directionality strongly limited the ability of the mass mixing models to general accurate fits. Simply, there werenÕt enough data points to produce reliable statistics for the gravitational potential and mass density parameters. 

\section{Future Study}\label{future}
\indent{}The pn-CCD has yet to be tested in this method, though in previous study, itÕs ability to adequately deal with background noise was much more limited than the MOS instruments. However, in the effort to improve statistical quality of results, its use in this form of analysis will be revisited. 

\indent{}Using the partial annulus mode described in \S\ref{annulusalgo}, the angular extent of the spectrum derived can be shrunk considerably, while the radial size can also be reduced to at least a factor of 4 smaller. Considering just the reduction in radial size, this would potentially produce \texttt{NFWMASS} and \texttt{CLMASS} fits with significantly reduced error, providing potentially 4 or more times the amount of data points for the mixing models to fit. It is also possible to extend farther into the core of the cluster, where the quality of the X-Ray data is significantly higher, due to both the higher incident flux from this region, as well as the existence of two separate observations of the same region. This will permit the ability to greatly improve the statistics of the model fittings for the innermost regions of the cluster.

\indent{} While this study has been unable to produce results of great statistical value, it has been shown that, with some adjustments, this method of analysis could provide great insight into the composition, temperature, and matter distributions of "local" clusters. With this in mind, it is logical that the next step is to examine other clusters with a similar methodology. Furthermore, the automated procedure outlined in this paper was designed for the purpose of wide scale application. Study of clusters such as Coma, Virgo, and Ophiucus are expected to take place in the not-too-distant future.

\section{Acknowledgements}

\indent{}I would like to thank several mentors and colleagues who have made this research paper possible. Professor Melville Ulmer has guided me for the past two years, constantly providing insight and motivation to move forward so that I could complete this paper. Professor Ulmer provided significant moral support and stress relief owls, as well as constantly being around to check the theoretical implications of my results. 

\indent{}I would like to that Matthew Wampler-Doty, for his expert assistance throughout the project. His mentorship -- essentially, teaching me how to program in multiple languages over the course of six months -- has been invaluable. Considerable effort and time was spent creating the backbone of the rotation matrix and spherical trigonometry, as well as advising on the creating of the automation framework. Finally, his assistance with some of the formatting and graphics for this paper was indispensable.  

\indent{}Paul Nulsen and Steve Snowden, creators of the \texttt{CLMASS} and \texttt{ESAS} package were both incredibly helpful throughout the process of analyzing data and writing this paper. Both made themselves available over phone and email regularly and were glad to answer any questions and concerns I had. 

\indent{}Thanks must also be made to the committee of the Illinois Space Grant Consortium, for the grant received during the Summer of 2011, during which a significant amount of preparation and initial testing that was integral to the completion of this thesis.

\bibliographystyle{plain}
\bibliography{bib}

\end{document}